\documentclass[preprint,nofootinbib,amsmath,prd,aps,superscriptaddress,tightenlines,12pt]{revtex4}
\usepackage[hypertex]{hyperref}
\usepackage{graphicx}
\usepackage{epstopdf}
\usepackage{bm}
\usepackage{epsfig}
\usepackage{graphics}
\usepackage{xspace}
\usepackage{amsmath}

\preprint{ANL-HEP-PR-11-35}
\preprint{nuhep-th/11-16}

\def\OMIT#1{}

\newcommand{\nn}{\nonumber}

\newcommand{\bn}{{\bar n}}
\newcommand{\bea}{\begin{eqnarray}}
\newcommand{\eea}{\end{eqnarray}}

\newcommand{\gsim}{\mathrel{\rlap{\lower4pt\hbox{\hskip1pt$\sim$}}\raise1pt\hbox{$>$}}}

\newcommand{\be}{\begin{equation}}
\newcommand{\ee}{\end{equation}}

\begin{document}
\setlength\baselineskip{17pt}


\title{\bf  An Exclusive Soft Function for Drell-Yan at Next-to-Next-to-Leading Order}


\author{Ye Li}
\email[]{yeli2012@u.northwestern.edu}
\affiliation{Department of Physics \& Astronomy, Northwestern University, Evanston, IL 60208, USA}
\author{Sonny Mantry}
\email[]{mantry147@gmail.com}
\affiliation{University of Wisconsin, Madison, WI, 53706}
\author{Frank Petriello}
\email[]{f-petriello@northwestern.edu}
\affiliation{Department of Physics \& Astronomy, Northwestern University, Evanston, IL 60208, USA}
\affiliation{High Energy Physics Division, Argonne National Laboratory, Argonne, IL 60439, USA}



\newpage
\begin{abstract}
  \vspace*{0.3cm}
  
 We present next-to-next-to-leading-order (NNLO) results for an exclusive soft function that appears in a recently developed factorization theorem for transverse momentum distributions. The factorization theorem, derived using the Soft Collinear Effective Theory, involves both a soft function and unintegrated nucleon distribution functions fully differential in momentum coordinates.  The soft function is given by the vacuum matrix element of soft Wilson lines 
and is also fully differential in all components.  We give results and relevant technical details for the NNLO calculation of the soft function, including finite parts,  and derive the corresponding anomalous dimension.  These results are necessary for achieving low transverse momentum resummation at next-to-next-to-leading-logarithmic accuracy in this effective field theory approach with unintegrated distribution functions.

\end{abstract}

\maketitle

\newpage
\section{Introduction}

Factorization plays a crucial role in collider physics by allowing for a more predictive framework through the separation of perturbative and non-perturbative effects. For processes such as fully inclusive Drell-Yan production of lepton pairs, factorization expresses the theoretical prediction as the convolution of a perturbatively calculable partonic cross-section with universal non-perturbative parton distribution functions (PDFs). Large logarithms of the hard and soft scales are resummed by evolving the PDFs from $\Lambda_{QCD}$ up to the hard scale, of order the invariant mass of the final-state leptons, via the DGLAP renormalization-group equations. For more exclusive processes, restrictions on the final state can introduce new  functions associated with intermediate momentum scales. In such cases, additional resummation is required and often new non-perturbative functions beyond the standard PDFs can arise. 

 An important example of such an observable is the low transverse momentum ($p_T$)  distribution of electroweak gauge bosons and the Higgs boson.  It plays an important role in the precision measurement of the $W$-boson mass, Higgs boson searches,  tests of perturbative Quantum Chromodynamics (QCD), and in probing non-perturbative transverse momentum dynamics in the nucleon. For perturbative values of $p_T$, three distinct scales appear in this problem, $M \gg p_T \gg \Lambda_{QCD}$, leading to a more intricate factorization formula.  Large logarithms of $M / p_T$ and $p_T/\Lambda_{QCD}$ must be resummed. In the non-perturbative region where $p_T\sim \Lambda_{QCD}$, new non-perturbative structures that probe transverse momentum dynamics in the nucleon appear. 
 
 Low $p_T$ distributions have been extensively studied in the 
traditional QCD literature~\cite{Dokshitzer:1978yd,Parisi:1979se,Curci:1979bg,Collins:1981uk,Collins:1984kg,Kauffman:1991jt,Yuan:1991we,Ellis:1997ii,Berger:2002ut,Bozzi:2003jy,
Bozzi:2005wk, Bozzi:2010xn, Aybat:2011zv} and have also been explored ~\cite{Gao:2005iu, Idilbi:2005er} in the context of the  Soft-Collinear Effective Theory 
(SCET)~\cite{Bauer:2000yr, Bauer:2001yt, Bauer:2002nz}. Recently, a new approach, based on SCET and on fully unintegrated nucleon distribution functions, was developed in Refs.~\cite{Mantry:2009qz, Mantry:2010mk,Mantry:2010bi}, resulting in a new  factorization and resummation theorem for low-$p_T$ distributions.  This approach allows one to predict the perturbative $p_T$ distribution entirely in terms
of perturbatively calculable functions and the standard PDFs, avoiding the difficulties in matching the low and high $p_T$ regions~\cite{Qiu:2000hf,Bozzi:2005wk} associated with treating the Landau pole in traditional approaches. The factorization theorem in this new approach takes the schematic form 
\bea
\label{fac-sketch}
\frac{d^2\sigma}{dp_T^2 \> dY} &\sim & H \otimes  \tilde{B}_n \otimes \tilde{B}_{\bn} \otimes {\cal S}^{-1},
\eea
where $H$ denotes a hard function, $\tilde{B}_{n,\bn}$ are Impact-parameter Beam Functions (iBFs),
and $ {\cal S}^{-1}$ is the Inverse Soft Function (iSF).  All objects have well-defined operator definitions, as shown in Refs.~\cite{Mantry:2009qz, Mantry:2010mk}.  The hard function $H$ is perturbatively calculable and encodes hard physics of the Drell-Yan production vertex. The iBFs are fully unintegrated nucleon distribution functions and the soft function ${\cal S}$ is given by the vacuum matrix element of soft Wilson lines. The iSF ${\cal S}^{-1}$ appears instead of ${\cal S}$  because of zero-bin subtractions~\cite{Manohar:2006nz, Lee:2006nr, Idilbi:2007ff}, necessary to avoid double counting the soft region, as explained in Refs.~\cite{Mantry:2009qz, Mantry:2010mk}. Similar soft-subtractions~\cite{Collins:1999dz, Hautmann:2007uw} appear in the formalism based on transverse-momentum dependent PDFs (TMDPDFs)~\cite{Collins:1999dz, Belitsky:2002sm, Hautmann:2007uw, Collins:1992kk, Ji:2002aa, Collins:2003fm, Ji:2004wu,  Cherednikov:2007tw, Cherednikov:2008ua,Cherednikov:2009wk, Aybat:2011zv}. For non-perturbative values of $p_T$, the iBFs and the iSF are non-perturbative functions that encode the physics of non-perturbative $p_T$ emissions and  transverse momentum dynamics in the initial state nucleons. For perturbative values of $p_T$, the iBFs describe the evolution and shattering of the initial state nucleon into an initial state beam-jet of high energy $p_T$ radiation. Analogous beam functions were first shown to arise in other contexts~\cite{Fleming:2006cd, Stewart:2009yx} and correspond to a special case of the iBF.  The iSF describes the emission of low energy $p_T$ radiation from the initial states. In the standard approaches, rapidity divergences arise in perturbative computations of the TMDPDFs that are not regulated in dimensional regularization and are instead regulated with additional external regulators. In contrast, the iBF, which is more differential in momentum coordinates than the TMDPDF, can be computed in standard dimensional regularization with rapidity divergences  regulated  by the physical kinematics of the process. 

  For perturbative values of $p_T$, the iBFs can be perturbatively matched onto the standard PDFs, thus factorizing the non-perturbative dynamics of the initial state nucleon from the perturbative $p_T$ emissions. In this case, the factorization theorem for the cross-section, differential in the $p_T$ and rapidity ($Y$) of the electroweak gauge boson, takes the form
\bea
\label{beam-soft-scales}
\frac{d^2\sigma}{dp_T^2\> dY}&=& \frac{\pi^2 }{N_c}   \int_0^1 dx_1 \int_0^1 dx_2\int_{x_1}^1 \frac{dx_1'}{x_1'} \int_{x_2}^1 \frac{dx_2'}{x_2'}   \nn \\
&\times&   H_Z^{q}(x_1x_2Q^2,\mu_Q;\mu_T) \>{\cal G}^{qrs}(x_1,x_2,x_1',x_2',p_T,Y,\mu_T) f_r(x_1',\mu_T)  f_s(x_2',\mu_T),\nn \\  
\eea
where $Q$ denotes the hadronic center of mass energy and the Transverse Momentum Function (TMF) function ${\cal G}^{qrs}$ is given by
\bea
\label{intro-DY-2}
 &&{\cal G}^{qrs}(x_1,x_2,x_1',x_2',p_T,Y,\mu_T)=  \int \frac{d^2b_\perp}{(2\pi)^2} J_0\big [b_\perp p_T\big ]\>\int dt_n^+ dt_\bn^- \> {\cal I}_{n;q r}(\frac{x_1}{x_1'}, t_n^+,b_\perp,\mu_T)\>\nn \\
&\times&{\cal I}_{\bn;\bar{q} s}(\frac{x_2}{x_2'}, t_\bn^-,b_\perp,\mu_T) {\cal S}^{-1}(x_1 Q-e^{Y}\sqrt{\text{p}_T^2+M^2}-\frac{t_\bn^-}{x_2 Q}, x_2 Q-e^{-Y}\sqrt{\text{p}_T^2+M^2}- \frac{t_n^+}{x_1 Q},b_\perp,\mu_T).\nn \\
\eea
The functions ${\cal I}_{n;q r}$ and ${\cal I}_{\bn;q s}$ are Wilson coefficients that arise from the perturbative matching of the iBFs onto the PDFs and are given by the finite part of the iBF computed in pure dimensional regularization.  The leading-order (LO) and next-to-leading-order (NLO) expressions for the iBFs and the iSF were computed in Ref.~\cite{Mantry:2010mk} and were used to calculate the next-to-leading-log (NLL) perturbative $p_T$ spectrum for the $Z$-boson. A resummation at the next-to-next-leading-log (NNLL) level requires a computation of the iBFs and the iSF at next-next-to-leading-order (NNLO) in perturbation theory. 

In this paper, we take the first step towards achieving a NNLL resummation of the Drell-Yan $p_T$-spectrum, using the effective field 
theory approach with unintegrated distribution functions, by computing the soft function ${\cal S}$ that appears in 
Eqs.~(\ref{beam-soft-scales}) and (\ref{intro-DY-2}) at NNLO. We perform several consistency checks on our calculation.  Both the result 
for this exclusive soft function, and the techniques used in deriving it, should be of use in other investigations of resummation to high 
accuracy within effective field theory. Recently~\cite{Hoang:2008fs,Kelley:2011ng,Monni:2011gb,Hornig:2011iu}, two-loop results for
a related soft function that appears in thrust distributions~\cite{Catani:1992ua, Korchemsky:1994is, Kidonakis:1998bk, Korchemsky:2000kp, Schwartz:2007ib, Hoang:2007vb, Bauer:2002ie, Bauer:2003di, Fleming:2007xt,Fleming:2007qr} of $e^+e^-$ collisions were given, demonstrating the the arising need for studying higher-order corrections to multi-scale objects appearing in factorization theorems.   Our paper is organized as follows.  We formulate the problem and introduce our notation in Section~\ref{sec:setup}.  We present both the techniques for and results of our calculation, including a comparison with known results, in Section~\ref{sec:calc}.  Finally, 
we conclude in Section~\ref{sec:conc}.

\section{Notation and NLO Results \label{sec:setup}}

The operator definition of the soft function in position space is given by
\begin{equation}
\label{position-space-soft-intro}
S(b,\mu) = \frac{1}{N_c}{\rm Tr} \langle 0 | \bar{T} [ S_n^{\dagger} S_{\bn} ] (b) \;T [ S_{\bn}^{\dagger} S_n ] (0) | 0 \rangle ,
\end{equation}
where $N_c=3$ denotes the number of colors. $S_{n,\bn}$ denote soft Wilson lines along the $n^\mu, \bn^\mu$ directions respectively and are defined as
\begin{equation}
S_{n} = P \, {\rm exp} \left[ ig \int_{-\infty}^0 ds \, n \cdot A_s (x+sn) \right], \qquad S_{n}^\dagger = \bar{P} \, {\rm exp} \left[ -ig \int_{-\infty}^0 ds \, n \cdot A_s (x+sn) \right],
\end{equation}
with analogous definitions for $S_{\bn}$ and $S_{\bn}^\dagger$. The four-vectors $n^\mu,\bn^\mu$ are light-like  and satisfy $n\cdot \bn=2$. The symbols $P, \bar{P}$ denote path-ordering and anti-path-ordering respectively, and similarly $T$, $\bar{T}$ denote time-ordering and anti-time-ordering.   

One can define a hybrid soft function with light-cone momentum coordinates and position space impact-parameter coordinates as the Fourier transform of $S(b,\mu)$ with respect to the light-cone coordinates as  %
\begin{equation}
\label{hybrid-space-soft}
{\cal S} (q^-, q^+, b_{\perp},\mu) = \int \frac{db^+ db^-}{16\pi^2} e^{iq^- b^+/2}e^{iq^+ b^-/2} \, S(b,\mu),
\end{equation}
It is this hybrid soft function that appears in the factorization theorem in Eq.~(\ref{beam-soft-scales}). Similarly, the full momentum space soft function is defined as
\begin{equation}
{\cal S}(q,\mu) = 2\int \frac{d^{d-2} b_{\perp}}{(2\pi)^{d-2}} \>e^{i \vec{q}_{\perp} \cdot \vec{b}_{\perp}} \>{\cal S} (q^-, q^+, b_{\perp},\mu)
\end{equation}
or equivalently, is related to the full position space soft function in the standard manner
\begin{equation}
\label{soft-func-to-calc}
{\cal S}(q,\mu) =   \int \frac{d^{d} b}{(2\pi)^{d}} \>e^{i q \cdot b}\>S(b,\mu).
\end{equation}
We will present results for the exclusive NNLO  position-space soft function $S(b,\mu)$ and the hybrid impact-parameter space soft function ${\cal S}(q^-,q^+,b_\perp)$ that appears directly in the factorization theorem of Ref.~\cite{Mantry:2010mk}.  The momentum-space soft function ${\cal S}(q)$ of Eq.~(\ref{soft-func-to-calc}) is used in intermediate stages of the calculations, and is simple to derive using the presented formulae.  We note that the position-space soft function of Eq.~(\ref{position-space-soft-intro}) by definition is equal to a gauge invariant soft function but evaluated in covariant gauges. In non-covariant or singular gauges such as the light-cone gauge, additional transverse gauge links are required in the definition of the soft function~\cite{Idilbi:2010im,GarciaEchevarria:2011md}. These transverse gauge links are unity in covariant gauges and thus do not appear in the definition of Eq.~(\ref{position-space-soft-intro}). Similar arguments also appear in Ref.~\cite{Belitsky:2002sm, Collins:2003fm} where the TMDPDFs were calculated in non-singular gauges. We restrict our analysis here to covariant gauges so that gauge invariance is fully respected.

\subsection{Renormalization}
Here we discuss the renormalization conventions for the hybrid impact-parameter space soft function ${\cal S}(q^-,q^+,b_\perp,\mu)$ that appears in the factorization theorem of Eq.~(\ref{beam-soft-scales}) and for the position-space soft function $S(b,\mu)$ of Eq.~(\ref{position-space-soft-intro}).  We regulate infrared and ultraviolet divergences using pure dimensional regularization with $d=4-2\epsilon$ and work in the $\overline{\text{MS}}$ renormalization scheme.

The renormalized hybrid-impact-parameter space soft function ${\cal S}(q^-, q^+, b_\perp, \mu) $ is related to the bare function
${\cal S}_b(q^-, q^+, b_\perp) $ as
\bea
\label{soft-renorm}
{\cal S}(q^-, q^+, b_\perp, \mu) &=& \int d\omega_1 \int d\omega_2 \>\>Z_S^{-1} (q^--\omega_1,q^+-\omega_2,\mu) \> {\cal S}_b (\omega_1,\omega_2,b_\perp), \nn \\
\eea
where $Z_S(q^--\omega_1,q^+-\omega_2,\mu)$ is the ultraviolet renormalization constant with an expansion around the $\epsilon \to 0$ limit given by
\bea
\label{Zsepsilon}
Z_S (\omega_1,\omega_2,\mu) &=& \delta(\omega_1)\delta(\omega_2) + \sum_{k=1}^\infty \frac{1}{\epsilon^k} Z_{S,k}(\omega_1,\omega_2,\mu),\nn \\
Z_S^{-1} (\omega_1,\omega_2,\mu) &=& \delta(\omega_1)\delta(\omega_2) + \sum_{k=1}^\infty \frac{1}{\epsilon^k} \bar{Z}_{S,k}(\omega_1,\omega_2,\mu). \nn \\
\eea
The $Z_{S,k}$ and $\bar{Z}_{S,k}$ can be related to each other from the condition 
\bea
\int d\omega_1' \int d\omega_2' \>\>Z_S^{-1} (\omega_1-\omega_1',\omega_2-\omega_2',\mu)Z_S(\omega_1'-\omega_1'',\omega_2'-\omega_2'',\mu) = \delta(\omega_1-\omega_1'') \delta(\omega_2-\omega_2''). \nn \\
\eea
The bare and renormalized strong couplings $\alpha_s^b$ and $\alpha_s$ respectively are related by the renormalization constant $Z_g$ as
\bea
\label{alpha-renorm}
\alpha_s^b &=& \mu_0^{2\epsilon} Z_g^2 \alpha_s(\mu), \qquad \!\! Z_g = 1 + \sum _{j=1}^\infty \Big [  \frac{\alpha_s(\mu)}{\pi} \Big ]^j z_{gj}, \qquad \!\! \mu_0^2 =\frac{\mu^2}{4\pi e^{-\gamma_E}}.
 \eea
 where $z_{g1} = (N_F/12-11C_A/24)/\epsilon$ at NLO.

 The renormalization group evolution of the soft function is determined by the equation
 \bea
 \mu \frac{d}{d\mu} {\cal S}(q^-,q^+,b_\perp,\mu) &=& \int d\omega_1\int d\omega_2 \>\gamma_{{\cal S}}(q^--\omega_1,q^+-\omega_2,\mu)\> {\cal S}(\omega_1,\omega_2,b_\perp,\mu), \nn \\
 \eea
where the anomalous dimension is defined as
\bea
\label{anom-dim-5}
\gamma_{{\cal S}}(\omega_1,\omega_2,\mu) &=& -\int d\omega_1'\int d\omega_2' \>Z_S^{-1} (\omega_1-\omega_1',\omega_2-\omega_2',\mu) \>\mu \frac{d}{d\mu} Z_S(\omega_1',\omega_2',\mu). \nn \\
\eea
From the finiteness of the anomalous dimension it can be shown that at any order in perturbation theory it is given by
 \bea
 \label{anom-dim}
\gamma_{{\cal S}}(\omega_1,\omega_2,\mu) &=& - 2\alpha_s \frac{\partial }{\partial \alpha_s} \bar{Z}_{S,1}(\omega_1,\omega_2,\mu) = 2\alpha_s \frac{\partial }{\partial \alpha_s} Z_{S,1}(\omega_1,\omega_2,\mu).
\eea
The $\alpha_s$ expansion of the anomalous dimension is defined as
\bea
\gamma_{{\cal S}}(\omega_1,\omega_2,\mu) &=& \sum_{n=1}^\infty \Big [\frac{\alpha_s(\mu)}{\pi} \Big ]^{(n)} \gamma_S^{(n)}(\omega_1,\omega_2,\mu)
\eea
Including the $\alpha_s$ expansion for each pole term in Eq.~(\ref{Zsepsilon}), we can write an expansion for $Z_S (\omega_1,\omega_2,\mu)$ and $Z_S^{-1} (\omega_1,\omega_2,\mu)$ in $\alpha_s$ and $\epsilon$ as
\bea
Z_S (\omega_1,\omega_2,\mu) &=& \delta(\omega_1)\delta(\omega_2) + \sum_{k=1}^\infty \sum_{j=1}^\infty \frac{1}{\epsilon^k}\Big [ \frac{\alpha_s(\mu)}{\pi}\Big ]^j Z_{S,k}^{(j)}(\omega_1,\omega_2,\mu). \nn \\
Z_S^{-1} (\omega_1,\omega_2,\mu) &=& \delta(\omega_1)\delta(\omega_2) + \sum_{k=1}^\infty \sum_{j=1}^\infty \frac{1}{\epsilon^k}\Big [ \frac{\alpha_s(\mu)}{\pi}\Big ]^j \bar{Z}_{S,k}^{(j)}(\omega_1,\omega_2,\mu). \nn \\
\eea
The renormalized and bare soft functions have  perturbative expansions given by
 \bea
 \label{soft-expand}
{\cal S}(q^-,q^+, b_\perp,\mu) &=& \sum_{j=0}^\infty \Big [\frac{\alpha_s(\mu)}{\pi} \Big ]^j {\cal S}^{(j)} (q^-,q^+, b_\perp,\mu), \nn \\
 {\cal S}_b(q^-,q^+, b_\perp) &=& \sum_{j=0}^\infty \Big [\frac{\alpha_s^b}{\pi} \Big ]^j {\cal S}_{b}^{(j)} (q^-,q^+, b_\perp ).\nn \\
 \eea
The renormalized soft function can be obtained at each order in perturbation theory by using Eq.~(\ref{alpha-renorm}) to write the bare coupling in terms of the renormalized coupling and then equating powers of $\alpha_s$ in Eq.~(\ref{soft-renorm}). The resulting consistency conditions at LO, NLO, and NNLO are given by
\bea
\label{renorm-consist}
{\cal S}^{(0)}(q^-,q^+, b_\perp,\mu) &=& {\cal S}_{b}^{(0)}(q^-,q^+, b_\perp) = \delta(q^-)\delta(q^+), \nn \\
{\cal S}^{(1)}(q^-,q^+, b_\perp,\mu) &=& \mu_0^{2\epsilon} {\cal S}_{b}^{(1)}(q^-,q^+, b_\perp)+\sum_{k=1}^\infty \frac{1}{\epsilon^k} \bar{Z}^{(1)}_{s,k} (q^-,q^+,\mu) , \nn \\
{\cal S}^{(2)}(q^-,q^+, b_\perp,\mu) &=&  2 z_{g1} \mu_0^{2\epsilon} {\cal S}_{b}^{(1)}(q^-,q^+, b_\perp)+  \mu_0^{4\epsilon} {\cal S}_{b}^{(2)}(q^-,q^+, b_\perp) \nn \\
&+& \sum_{k=1}^\infty \frac{1}{\epsilon^k}\bar{Z}^{(2)}_{S,k} (q^-,q^+,\mu) \nn \\
&+& \mu_0^{2\epsilon} \int d\omega_1 \int d\omega_2\sum_{k=1}^\infty \frac{1}{\epsilon^k}\bar{Z}^{(1)}_{S,k} (q^--\omega_1,q^+-\omega_2,\mu) \>{\cal S}_{b}^{(1)}(\omega_1,\omega_2,b_\perp). \nn \\
\eea
In the $\overline{\text{MS}}$ scheme the renormalization constants  $\bar{Z}^{(1)}_{S,k}$ and $\bar{Z}^{(2)}_{S,k}$ are determined by requiring a cancellation of all pole terms of the RHS above  in order to yield a finite result for the renormalized soft function.

A similar analysis can be done for the position-space soft function, and we outline the main features  below to establish notation. In position space, the  renormalized and bare soft functions $S(b,\mu)$ and $S_b(b)$ respectively are related by a multiplicative renormalization constant so that
\bea
\label{bare-renorm-soft-b}
S(b,\mu) &=& Z_s^{-1}(b,\mu) S_b(b).
\eea
The anomalous dimension of $S(b,\mu)$ is defined as
\bea
\mu \frac{d}{d\mu} S(b,\mu) = \gamma_{S}(b^+,b^-,\mu)\> S(b,\mu),\qquad \gamma_{{\cal S}}(b,\mu) = -Z_S^{-1} (b^+,b^-,\mu) \>\mu \frac{d}{d\mu} Z_S(b^+,b^-,\mu), \nn \\
\eea
with an expansion in $\alpha_s$ given by
\bea
\gamma_{{\cal S}}(b^+,b^-,\mu) &=& \sum_{n=1}^\infty \Big [\frac{\alpha_s(\mu)}{\pi} \Big ]^{(n)} \gamma_S^{(n)}(b^+,b^-,\mu).
\eea
The expansion in $\alpha_s$ and $\epsilon $ of the renormalization constants are defined as
\bea
\label{renorm-const-pos}
   Z_s(b^+,b^-,\mu) &=& 1 + \sum_{k=1}^\infty \sum_{j=1}^\infty \frac{1}{\epsilon^k}\Big [ \frac{\alpha_s(\mu)}{\pi}\Big ]^j Z_{S,k}^{(j)}(b^+,b^-,\mu), \nn \\
     Z_s^{-1}(b^+,b^-,\mu) &=& 1 + \sum_{k=1}^\infty \sum_{j=1}^\infty \frac{1}{\epsilon^k}\Big [ \frac{\alpha_s(\mu)}{\pi}\Big ]^j \bar{Z}_{S,k}^{(j)}(b^+,b^-,\mu), \nn \\
\eea
where once again the $Z_{S,k}^{(j)}(b^+,b^-,\mu)$ and $\bar{Z}_{S,k}^{(j)}(b^+,b^-,\mu)$ coefficients will be related to each other by the condition $ Z_s^{-1}(b^+,b^-,\mu)  Z_s(b^+,b^-,\mu)=1$. The perturbative expansion for the bare and renormalized position-space soft functions are defined as
 \bea
S(b,\mu) &=& \sum_{j=0}^\infty \Big [\frac{\alpha_s(\mu)}{\pi} \Big ]^j S^{(j)} (b,\mu), \qquad
 S_b(b) = \sum_{j=0}^\infty \Big [\frac{\alpha_s^b}{\pi} \Big ]^j S_{b}^{(j)} (b ),\nn \\
 \eea
 and the joint expansion in $\alpha_s^b$ and $\epsilon$ of the bare soft function is defined as
 \bea
 \label{pos-space-expan}
S_b(b)=1+ \sum_{k=1}^\infty \sum_{j=1}^\infty \frac{1}{\epsilon^k}\Big [ \frac{\alpha_s^b}{\pi}\Big ]^j S_{b,k}^{(j)}(b).\nn \\
\eea
Following the same procedure as in the case of the hybrid impact-parameter space soft function ${\cal S}(q^-,q^+,b_\perp,\mu)$, consistency  conditions analogous to Eq.~(\ref{renorm-consist}) can be derived for the position-space  soft functions $S(b,\mu)$ as well.

\subsection{NLO Soft Function}

We calculate the soft function by inserting a complete set of  states using the identity $1=  \sum | X_s \rangle \langle X_s |$  in Eq.~(\ref{position-space-soft}) to get

\begin{equation}
\label{position-space-soft}
S(b,\mu) = \frac{1}{N_c}\sum_{X_s}{\rm Tr} \langle 0 | \bar{T} [ S_n^{\dagger} S_{\bn} ] (b) | X_s \rangle \langle X_s | \;T [ S_{\bn}^{\dagger} S_n ] (0) | 0 \rangle ,
\end{equation}
and then compute the resulting product of matrix elements  in each term above.
 At LO in QCD perturbation theory,  only the vacuum state $| X_s \rangle = |0 \rangle$ contributes.  At NLO, both virtual corrections to the vacuum insertion and the single gluon final state $| X_s \rangle = |\epsilon_g^A(k) \rangle$ contribute.  In pure dimensional regularization the virtual graphs are scaleless and a non-vanishing contribution arises only from the real emission of a gluon into the final state.

From the results of Ref.~\cite{Mantry:2010mk}, the LO and NLO  terms of the bare soft function ${\cal S}_b(q^-,q^+,b_\perp)$ of Eq.~(\ref{soft-expand}) are known to be
\begin{eqnarray}
\label{soft-NLO}
{\cal S}_b^{(0)}(q^+,q^-,b_{\perp}) &=& \delta ( q^{+})\delta ( q^{-}),\nonumber \\
{\cal S}_b^{(1)}(q^+,q^-,b_{\perp}) &=& C_F \frac{(4\pi)^{\epsilon}}{\Gamma(1-\epsilon)} \left(q^+ q^- \right)^{-1-\epsilon} {_{0}F_{1}}\left( 1-\epsilon ; -\frac{b_{\perp}^2 q^+ q^-}{4} \right).
\end{eqnarray} 
We note that $q^{\pm}$ are constrained to be positive quantities.  Using the NLO consistency condition corresponding to the second equation in Eq.~(\ref{renorm-consist}) and  using  Eq.~(\ref{anom-dim}), the anomalous dimension at one loop $\gamma_{{\cal S}}^{(1)}$ is given by
\bea
\label{soft-NLO-1}
\gamma_{{\cal S}}^{(1)}(q^-,q^+,\mu) = -\frac{2\alpha_s}{\pi} C_F \bigg [\>\frac{1}{\mu} \bigg [\frac{\mu}{q^-}  \bigg ]_+ \delta(q^+) + \frac{1}{\mu} \bigg [\frac{\mu}{q^+}  \bigg ]_+ \delta(q^-)\> \bigg ].
\eea
 We note that we have also performed this NLO calculation using an off-shell regulator for infrared divergences and have obtained the same results as above.

In full momentum space, the LO and NLO coefficients of the bare soft function ${\cal S}_b(q,\mu)$ are given by
\begin{eqnarray}
\label{low-order-mom}
{\cal S}_b^{(0)}(q) &=& \delta^{(d)}(q),\nonumber \\
{\cal S}_b^{(1)}(q) &=& \frac{2 C_F}{\pi^{1-\epsilon}} (4\pi)^{\epsilon} \frac{\delta_{+}(q^2)}{q^+ q^-},
\end{eqnarray}
where the $+$ subscript on the delta function denotes that only the positive energy solution is taken.  It is straightforward to check that the Fourier transform of these coefficients with respect to $\vec{q}_\perp$ give the coefficients of Eq.~(\ref{soft-NLO}) in hybrid-impact-parameter space.

Finally, the soft function in full position space is obtained by Fourier transforming the result in Eq.~(\ref{low-order-mom}) to get
\begin{eqnarray}
S_b^{(0)}(b) &=& 1,\nonumber \\
S_b^{(1)}(b) &=& C_F \frac{\Gamma(1-\epsilon)}{\epsilon^2} e^{-\epsilon \gamma_E} \mu_0^{-2\epsilon} L^{\epsilon}\> \>{_{2}F_{1}}\left(-\epsilon,-\epsilon;1-\epsilon ; \frac{b_{\perp}^2}{b^{+}b^{-}} \right).
\end{eqnarray}
where we have defined
\bea
L \equiv -b^{+}b^{-} \mu^2 e^{2\gamma_E} /4.
\eea
Upon taking only a time-like component $b \to b^{0}$, this is in agreement with Ref.~\cite{Belitsky:1998tc} (we note that $\mu_0$ must be set to unity when 
making the comparison of our bare results with Ref.~\cite{Belitsky:1998tc}, as our $S_b^{(j)}$ are defined as coefficients of the bare coupling constant 
rather than the renormalized one).  The one loop anomalous dimension $\gamma_{S}^{(1)}$ for the position space soft function $S(b,\mu)$ is given 
by
\bea
\gamma_{S}^{(1)} (b,\mu) &=& \frac{2\alpha_s}{\pi} C_F \ln L.
\eea

\section{The soft function at NNLO  \label{sec:calc}}

\begin{figure}
\includegraphics[scale=0.9]{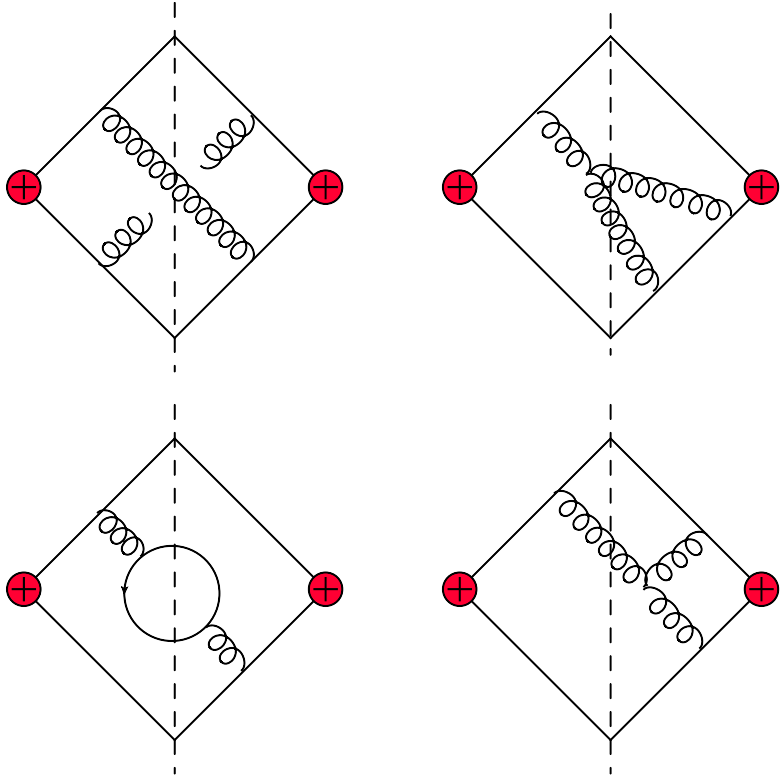}
\caption{Example diagrams contributing the the soft function at next-to-next-to-leading order.  From the top left diagram and proceeding in a clockwise fashion, the diagrams are respectively contributing to the terms
${\cal S}^{(2)}_{RR,C_F^2}$, ${\cal S}^{(2)}_{RR,C_F CA}$, ${\cal S}^{(2)}_{RV,C_FC_A}$ and ${\cal S}^{(2)}_{RR,C_F N_F}$.}
\label{soft_NNLO}
\end{figure}

For the calculation of the soft function at NNLO, we again start with  Eq.~(\ref{position-space-soft}) for the position-space soft function. Contributions will arise from two-loop virtual corrections to the vacuum state ($S^{(2)}_{VV} $), one-loop virtual corrections to single-gluon emission ($S^{(2)}_{RV} $), and the  real emission of two gluons or a quark-anti-quark pair in the final state ($S^{(2)}_{RR} $) so that we can write
\begin{equation}
S^{(2)}(b,\mu) = S^{(2)}_{VV}(b,\mu) +S^{(2)}_{RV}(b,\mu) + S^{(2)}_{RR}(b,\mu).
\end{equation}
In pure dimensional regulation, the purely virtual contribution $S^{(2)}_{VV} $ is scaleless and vanishes and will not be studied in further detail.
Representative diagrams for the remaining non-vanishing contributions
are shown in Fig.~\ref{soft_NNLO}, where the dark solid lines represent the soft Wilson lines and the dashed line indicates the insertion of states that puts particles on their mass shells. These contributions can be further decomposed according to their color structure so that we can write
\begin{equation}
\label{soft-color}
S^{(2)}(b,\mu) = S^{(2)}_{VV}(b,\mu) +S^{(2)}_{RR,C_F^2} + S^{(2)}_{RR,C_F C_A}+S^{(2)}_{RR,C_F N_F}+S^{(2)}_{RV,C_F C_A},
\end{equation}
where the subscripts $C_F^2$, $C_F C_A$, and $C_F N_F$ denote corresponding the color structures and
$N_F$ is the number of massless fermions.  We will consider each of these contributions separately, presenting both the final result and the relevant technical details.  We will first present results for the bare soft function and then discuss renormalization and the the extraction of the anomalous dimension.  As checks of our results we will use both the comparison with the $b \to b^{0}$ limit of Ref.~\cite{Belitsky:1998tc} and the constraint 
of non-abelian exponentiation~\cite{Gatheral:1983cz,Frenkel:1984pz}.  We will present results in both position-space and in the hybrid-impact-parameter space.

\subsection{The bare soft function}

We begin by first presenting the results for the various contributions to the bare NNLO soft function in position space. The NNLO calculations are first performed in momentum space and then converted to position space by inverting  Eq.~(\ref{soft-func-to-calc}).  The most difficult integrals occur in the double real-radiation graphs; the necessary integrals are presented in Appendix ~\ref{integrals}.   The results for these integrals, which typically contain hypergeometric functions, must be Fourier transformed to position space.  To do so, we first series expand the hypergeometric functions in their argument, perform the integrals for each term of the series, and then identify the resulting sum. The relevant integrals for this procedure are presented in the Appendix.

The results for the various non-vanishing contributions to the bare soft function $S_b(b,\mu)$,  in pure dimensional regularization are
\allowdisplaybreaks\begin{eqnarray}
S^{(2)}_{b,RR,C_F N_F}(b) &=& - C_F N_F \frac{\Gamma^2(1-\epsilon)}{{\epsilon}^3}
\frac{1-\epsilon}{8(1-2\epsilon)(3-2\epsilon)} e^{-2\epsilon\gamma_E} \mu_0^{-4\epsilon} L^{2\epsilon}
{\;_2F_1(-2\epsilon,-2\epsilon;1-2\epsilon;\frac{b_{\perp}^2}{b^{+}b^{-}})},\nonumber \\
S^{(2)}_{b,RR,C_F^2}(b) &=& {C_F}^{2} \frac{\Gamma^2(1-\epsilon)}{4{\epsilon}^4} e^{-2\epsilon\gamma_E} \mu_0^{-4\epsilon} L^{2\epsilon}
\left[  {\;_3F_2(-2\epsilon,-2\epsilon,-2\epsilon;1-\epsilon,1-3\epsilon;\frac{b_{\perp}^2}{b^{+}b^{-}})}
      \right. \nonumber \\ &+&\left. {\;_2F_1(-2\epsilon,-2\epsilon;1-3\epsilon;\frac{b_{\perp}^2}{b^{+}b^{-}})} \right]  + {\rm O}(\epsilon), \nn  \\      
S^{(2)}_{b,RR,C_F C_A}(b) &=& C_A C_F \frac{\Gamma^2(1-\epsilon)}{16{\epsilon}^4} e^{-2\epsilon\gamma_E} \mu_0^{-4\epsilon} L^{2\epsilon}
\left\{    \frac{(2-\epsilon)(3-\epsilon)}{(1-2\epsilon)(3-2\epsilon)}
           {\;_2F_1(-2\epsilon,-2\epsilon;1-2\epsilon;\frac{b_{\perp}^2}{b^{+}b^{-}})} \right. \nonumber \\
&+&       2 \frac{\Gamma^2(1-2\epsilon)}{\Gamma(1-3\epsilon)\Gamma(1-\epsilon)}
           {\;_3F_2(-\epsilon,-\epsilon,-\epsilon;1-\epsilon,1-3\epsilon;1)} \nonumber \\
           &\times& {\;_3F_2(-2\epsilon,-2\epsilon;1-2\epsilon;1-\epsilon,1-3\epsilon;\frac{b_{\perp}^2}{b^{+}b^{-}})} 
-\left. 2 {\;_2F_1(-2\epsilon,-2\epsilon;1-3\epsilon;\frac{b_{\perp}^2}{b^{+}b^{-}})} \right\} \nn \\
&+& {\rm O}(\epsilon), \nonumber \\      
S^{(2)}_{b,RV,C_FC_A}(b) &=& - C_A C_F \frac{\Gamma^2(1-\epsilon)}{8{\epsilon}^4} e^{-2\epsilon\gamma_E} \mu_0^{-4\epsilon} L^{2\epsilon}
\left[\Gamma(1-2\epsilon)\Gamma^2(1+\epsilon)\cos(\pi\epsilon)\right] \nn \\
&\times& {\;_2F_1(-2\epsilon,-2\epsilon;1-\epsilon;\frac{b_{\perp}^2}{b^{+}b^{-}})}. \nonumber \\
\end{eqnarray}
For simplicity of presentation, we have expanded some of the results above only up to the needed order in $\epsilon$.  It is straightforward to confirm that in the limit $b_{\perp} \to 0$, these expressions are identical to those obtained in Ref.~\cite{Belitsky:1998tc}.  In addition, the full position-space result is constrained by non-abelian exponentiation~\cite{Gatheral:1983cz,Frenkel:1984pz}, which requires the $C_F^2$ terms at NLO and NNLO  to obey the relation  $ S^{(2)}_{RR,C_F^2}(b) = \left[ S^{(1)}(b) \right]^2/2$.  This relation can 
be checked through the finite order in $\epsilon$ using the expansions given in Eqs.~(\ref{2F1exp1},~\ref{2F1exp2},~\ref{3F2exp}).  These two checks are strong indications of the correctness of our results.

We now present the results for the soft function in the hybrid impact-parameter space, that appears directly in the formalism of Refs.~\cite{Mantry:2009qz, Mantry:2010mk}.  These results can be obtained through the Fourier transform of the position-space results. As in the case of the position-space soft function, we separate the results into several components dictated by the color and cut structure, and obtain the following expressions:
\begin{eqnarray}
S^{(2)}_{b,RR,C_F N_F}(q^-, q^+, b_{\perp}) &=&- C_F N_F (4\pi)^{2\epsilon} \frac{\Gamma^2(1-\epsilon)}{\epsilon\Gamma^2(1-2\epsilon)}
\frac{1-\epsilon}{2(1-2\epsilon)(3-2\epsilon)} (q^{+}q^{-})^{-1-2\epsilon} \nonumber \\ &&\times {\;_0F_1(1-2\epsilon;-\frac{{b_{\perp}}^2 q^{+}q^{-}}{4})},\nn \\
S^{(2)}_{b,RR,C_F^2} (q^-, q^+, b_{\perp}) &=& - {C_F}^{2}(4\pi)^{2\epsilon} \frac{\Gamma^2(1-\epsilon)}{\epsilon^2\Gamma^2(1-2\epsilon)}
(q^{+}q^{-})^{-1-2\epsilon} \Big [ {\;_0F_1(1-3\epsilon;-\frac{{b_{\perp}}^2 q^{+}q^{-}}{4})} \nn \\
&+&  {\;_1F_2(-2\epsilon;1-\epsilon,1-3\epsilon;-\frac{{b_{\perp}}^2 q^{+}q^{-}}{4})} \Big ] + {\rm O}(\epsilon), \nn \\
S^{(2)}_{b,RR,C_F C_A} (q^-, q^+, b_{\perp}) &=& C_A C_F (4\pi)^{2\epsilon} \frac{\Gamma^2(1-\epsilon)}{\epsilon^2\Gamma^2(1-2\epsilon)} (q^{+}q^{-})^{-1-2\epsilon}
\left\{    \frac{(2-\epsilon)(3-\epsilon)}{4(1-2\epsilon)(3-2\epsilon)} \right. \nonumber \\
      &&\times     {\;_0F_1(1-2\epsilon;-\frac{{b_{\perp}}^2 q^{+}q^{-}}{4})}  + \frac{\Gamma^2(1-2\epsilon)}{2\Gamma(1-3\epsilon)\Gamma(1-\epsilon)} \nonumber \\
     &&\times      {\;_3F_2(-\epsilon,-\epsilon,-\epsilon;1-\epsilon,1-3\epsilon;1)}
           {\;_1F_2(1-2\epsilon;1-\epsilon,1-3\epsilon;-\frac{{b_{\perp}}^2 q^{+}q^{-}}{4})} \nonumber \\
&&\left. - \frac{1}{2} {\;_0F_1(1-3\epsilon;-\frac{{b_{\perp}}^2 q^{+}q^{-}}{4})} \right\} +{\cal O}(\epsilon), \nn \\
S^{(2)}_{b,RV,C_FC_A} (q^-, q^+, b_{\perp}) &=& - C_A C_F (4\pi)^{2\epsilon} \frac{\Gamma^2(1-\epsilon)}{\Gamma^2(1-2\epsilon)}
\frac{\Gamma(1-2\epsilon)\Gamma^2(1+\epsilon)\cos(\pi\epsilon)}{2{\epsilon}^2} (q^{+}q^{-})^{-1-2\epsilon} \nonumber \\
&&\times {\;_0F_1(1-\epsilon;-\frac{{b_{\perp}}^2 q^{+}q^{-}}{4})}. \nn \\
\end{eqnarray}
The $C_F^2$ piece again satisfies non-abelian exponentiation, but with a convolution in momenta rather than a simple product as explained in the next section.

\subsection{Renormalization, exponentiation, and the finite soft function}

The soft function is constrained by non-abelian exponentiation~\cite{Gatheral:1983cz, Frenkel:1984pz} so that one can write the position-space soft function as 
\begin{equation}
\label{pos-expo-1}
S(b,\mu) =  \,{\rm exp}\big \{s(b,\mu)\big \}= \,{\rm exp} \left\{ \sum_{n=0} \left( \frac{\alpha_s}{\pi}\right)^n s^{(n)}(b,\mu) \right\},
\end{equation}
with appropriately defined $s(b,\mu)$ and $s^{(n)}(b,\mu)$. The bare and renormalized 
position-space soft functions are related as in Eq.~(\ref{bare-renorm-soft-b}). In exponentiated form, this relationship can be re-expressed as
\begin{eqnarray}
\label{pos-expo}
S(b,\mu) &=& \exp \left\{ \bar{z}_S(b^+,b^-,\mu)+s_b(b) \right\}, \nonumber \\
\eea
where
\bea
\label{renorm-3}
\bar{z}_S(b^+,b^-,\mu) =  \displaystyle\sum\limits_{n=1}^{\infty} \displaystyle\sum\limits_{m=1}^{2n} \left[\frac{\alpha_s(\mu)}{\pi}\right]^n \frac{\bar{z}^{(n)}_{S,m}}{\epsilon^m}, \qquad s_b(b)= \displaystyle\sum\limits_{n=1}^{\infty} \displaystyle\sum\limits_{m=-\infty}^{2n} \left[\frac{\alpha_{s}^b}{\pi}\right]^n \frac{s^{(n)}_{b,m}}{\epsilon^m} ,
\end{eqnarray}
with $\bar{z}_{S,m}^{(n)}=\bar{z}_{S,m}^{(n)}(b^+,b^-,\mu)$ and $s_{b,m}^{(n)}=s^{(n)}_{b,m}(b)$.  We have introduced a similar exponentiated form for the 
renormalization constant,
\begin{equation}
Z_s^{-1}(b^+,b^-,\mu) = \exp \left\{ \bar{z}_S(b^+,b^-,\mu) \right\}.
\end{equation}
The bare and renormalized strong couplings are related by Eq.~(\ref{alpha-renorm}). The non-abelian exponentiation theorem implies consistency relationships between the coefficients in Eqs.(\ref{renorm-const-pos}) and (\ref{pos-space-expan}) and  the coefficients $s_{b,m}^{(n)}$ and $\bar{z}_{S,m}^{(n)}$. 

The renormalization-scale independence of the bare soft function determines the anomalous dimension in terms of $\bar{z}_S$ as
\bea
\label{anom-3}
\mu \frac{d}{d\mu} S(b,\mu) = \gamma_S(b^+,b^-,\mu)S(b,\mu), \qquad \gamma_S(b^+,b^-,\mu) = \mu\frac{d}{d\mu} \bar{z}_S(b^+,b^-,\mu). \nn \\
\eea
The finiteness of the renormalized soft function $S(b,\mu)$ determines the renormalization constants in Eqs.~(\ref{pos-expo}), (\ref{renorm-3}). The LO and NLO anomalous dimension contributions are given by
\begin{eqnarray}
\gamma_S^{(1)} &=& 2 C_F \ln L, \nonumber \\
\gamma_S^{(2)} &=& C_F N_F \left[ -\frac{5}{9}\ln L-\frac{14}{27}+\frac{\pi^2}{36} \right] +
C_F C_A \left[ \left( \frac{67}{18}-\frac{\pi^2}{6} \right)\ln L + \frac{101}{27}-\frac{11\pi^2}{72}-\frac{7}{2}\zeta(3) \right] . \nonumber \\
\end{eqnarray}
By comparing the coefficients of the $\alpha_s$ expansion of the RHS of Eqs.~(\ref{pos-expo-1}) and (\ref{pos-expo}) after renormalization,  the NLO and NNLO terms in the exponent of the renormalized soft function $S(b,\mu)$ in Eq.~(\ref{pos-expo-1}) are determined to be
\begin{eqnarray}
s^{(1)}(b,\mu) &=& C_F \left\{ \frac{1}{2}\ln^2 L+\frac{\pi^2}{12}+{\rm Li}_2(\frac{b_{\perp}^2}{b^{+}b^{-}}) \right\}, \nonumber \\
s^{(2)}(b,\mu) &=& C_F N_F \left\{ - \frac{1}{36}\ln^3 L - \frac{5}{36}\ln^2 L
- \left(\frac{7}{27}+\frac{1}{6}{\rm Li}_2(\frac{b_{\perp}^2}{b^{+}b^{-}})\right)\ln L
-\frac{41}{162} - \frac {5\pi^2}{432} + \frac{\zeta(3)}{36} \right. \nonumber \\
&& \left. - \frac{5}{18}{\rm Li}_2(\frac{b_{\perp}^2}{b^{+}b^{-}}) - \frac{1}{6}{\rm Li}_3(\frac{b_{\perp}^2}{b^{+}b^{-}})
+ \frac{1}{6}{\rm S}_{1,2}(\frac{b_{\perp}^2}{b^{+}b^{-}}) \right\} + \nonumber \\
&& C_F C_A \left\{ {\frac {11}{72}} \ln^3 L + \left( {\frac {67}{72}}-\frac{\pi^2}{24} \right) \ln^2 L
+ \left( \frac {101}{54}-\frac{7}{4}\zeta(3)+\frac{11}{12} {\rm Li}_2(\frac{b_{\perp}^2}{b^{+}b^{-}}) \right) \ln L \right. \nonumber \\
&& +\frac{607}{324}+\frac{67\pi^2}{864}-\frac{11}{72}\zeta(3)-\frac{\pi^4}{48}
+ \left(\frac {67}{36}-\frac{\pi^2}{12}\right) {\rm Li}_2(\frac{b_{\perp}^2}{b^{+}b^{-}})+\frac{11}{12}{\rm Li}_3(\frac{b_{\perp}^2}{b^{+}b^{-}})
-\frac{1}{2}{\rm Li}_4(\frac{b_{\perp}^2}{b^{+}b^{-}}) \nonumber \\
&& \left. -\frac{11}{12}{\rm S}_{1,2}(\frac{b_{\perp}^2}{b^{+}b^{-}})
-{\rm S}_{1,3}(\frac{b_{\perp}^2}{b^{+}b^{-}})
-\frac{1}{4}{\rm Li}_2^2(\frac{b_{\perp}^2}{b^{+}b^{-}}) \right\}.
\end{eqnarray}
Here, ${\rm Li}_n(x)$ is the usual polylogarithmic function, ${\rm S}_{n,p}(x)$ is Nielsen's generalized polylogarithm, and we remind the reader that $L = -b^{+}b^{-} \mu^2 e^{2\gamma_E} /4$. 

Non-abelian exponentiation also holds in the hybrid-impact parameter space but with the multiplication operation replaced by a convolution in the light-cone momentum coordinates so that
\bea
\label{expo-3}
{\cal S}(q^-,q^+,b_\perp,\mu ) &=& {\cal S}^{(0)}(q^-,q^+,b_\perp)\otimes {\rm exp} \left\{s(q^-,q^+,b_\perp,\mu) \right\}, \nn \\
\eea
where the perturbative expansion of the exponent is defined as
\bea
\label{expo-4}
s(q^-,q^+,b_\perp,\mu) &=&  \sum_{n=1} \left( \frac{\alpha_S}{\pi}\right)^n {s}^{(n)} (q^-,q^+,b_\perp,\mu). \nn \\
\eea
It should be understood that the first term in the expansion of the exponential is $\delta(q^-)\delta(q^+)$.  
The first few terms obtained after expanding the exponential in Eq.~(\ref{expo-3}) take the form 
\begin{equation}
{\cal S}(q^-,q^+,b_\perp,\mu )= {\cal S}^{(0)} + {\cal S}^{(0)} \otimes \Big [ s + \frac{1}{2} \>s\otimes s +\cdots \Big ],
\end{equation}
where we have suppressed the arguments of the functions on the RHS.
The convolution product $\otimes$ in the light-cone momentum coordinates is defined as 
\begin{equation}
\label{conv-def}
f(q^-,q^+) \otimes g(q^-,q^+) \equiv \int d\omega_1 \int d\omega_2 \>f(q^--\omega_1,q^+-\omega_2) \>g(\omega_1,\omega_2).
\end{equation}
Similarly, the relationship between the renormalized and bare soft function in exponentiated form also involves convolution products and is given by
\begin{eqnarray}
{\cal S}(q^-,q^+,b_\perp,\mu) &=& {\cal S}^{(0)}_b(q^-,q^+,b_\perp)\otimes\exp \left\{ \bar{z}_S(q^-,q^+,\mu)+s_b(q^-,q^+,b_\perp) \right\}, \nonumber \\
\eea
where the expansion in $\alpha_s$ for the terms in the exponent are defined as
\bea
  \bar{z}_S(q^-,q^+,\mu)  = \displaystyle\sum\limits_{n=1}^{\infty} \displaystyle\sum\limits_{m=1}^{2n} \left(\frac{\alpha_S}{\pi}\right)^n \frac{z^{(n)}_{S,m}}{\epsilon^m},\qquad  s_b(q^-,q^+,b_\perp)= \displaystyle\sum\limits_{n=1}^{\infty} \displaystyle\sum\limits_{m=-\infty}^{2n} \left(\frac{\alpha_{0}}{\pi}\right)^n \frac{s^{(n)}_{b,m}}{\epsilon^m}, \nn \\
\end{eqnarray}
with  $\bar{z}_{S,m}^{(n)}=\bar{z}_{S,m}^{(n)}(q^-,q^+,\mu)$ and $s_{b,m}^{(n)}=s^{(n)}_{b,m}(q^-,q^+,b_\perp)$. The anomalous dimension and the renormalization constants again satisfy 
\bea
\label{anom-4}
\mu \frac{d}{d\mu} S(q^-,q^+,b_\perp,\mu) &=& \int d\omega_1 \int d\omega_2\> \gamma_S(q^--\omega_1,q^+-\omega_2,\mu)\>S(\omega_1,\omega_2,b_\perp, \mu), \nn \\\gamma_S(q^-,q^+,\mu) &=& \mu\frac{d}{d\mu} \bar{z}_S(q^-,q^+,\mu). \nn \\
\eea

For simplicity of notation, in the rest of this section we use the following definitions
\begin{eqnarray}
\label{aux-impact}
T(z) &=& J_0(z)\left(\ln(\frac{z}{2})+\gamma_E\right)-\frac{\pi}{2}Y_0(z), \nonumber \\ 
L_{0,0} &=& \delta(q^{-})\delta(q^{+}), \nonumber \\
L_{0,1} &=& \frac{1}{\mu} \left[\frac{\mu}{q^{+}}\right]_{+} \delta(q^{-}) + \frac{1}{\mu} \left[\frac{\mu}{q^{-}}\right]_{+} \delta(q^{+}), \nonumber \\
L_{0,2} &=& \frac{1}{\mu} \left[\frac{\ln(q^{+}/\mu)}{q^{+}/\mu}\right]_{+} \delta(q^{-}) 
          + \frac{1}{\mu} \left[\frac{\ln(q^{-}/\mu)}{q^{-}/\mu}\right]_{+} \delta(q^{+}), \nonumber \\
L_{0,3} &=& \frac{1}{\mu} \left[\frac{\ln^2(q^{+}/\mu)}{q^{+}/\mu}\right]_{+} \delta(q^{-}) 
          + \frac{1}{\mu} \left[\frac{\ln^2(q^{-}/\mu)}{q^{-}/\mu}\right]_{+} \delta(q^{+}), \nonumber \\
L_{1,1} &=& \frac{1}{\mu^2} \left[\frac{\mu}{q^{+}}\right]_{+} \left[\frac{\mu}{q^{-}}\right]_{+}, \nonumber \\
L_{1,2} &=& \frac{1}{\mu^2} \left[\frac{\ln(q^{+}/\mu)}{q^{+}/\mu}\right]_{+} \left[\frac{\mu}{q^{-}}\right]_{+}
          + \frac{1}{\mu^2} \left[\frac{\ln(q^{-}/\mu)}{q^{-}/\mu}\right]_{+} \left[\frac{\mu}{q^{+}}\right]_{+}.
\end{eqnarray}
We note that $T(0)=0$.  The LO and NLO anomalous dimension contributions in the hybrid-impact-parameter space are given by
\begin{eqnarray}
\gamma_S^{(1)}(q^-,q^+,\mu) &=& -2 C_F L_{0,1}, \nonumber \\
\gamma_S^{(2)}(q^-,q^+,\mu) &=& \left\{ C_F N_F \left[ -\left(\frac{14}{27}-\frac{\pi^2}{36}\right)L_{0,0} + \frac{5}{9}L_{0,1} \right] + \right. \nonumber \\
&& \left. C_F C_A \left[ \left(\frac{101}{27}-\frac{11\pi^2}{72}-\frac{7}{2}\zeta(3)\right) L_{0,0} 
- \left( \frac{67}{18}-\frac{\pi^2}{6} \right)L_{0,1} \right] \right\}. \nn \\
\end{eqnarray}
The same results for the anomalous dimension can be derived using Eqs.~(\ref{anom-dim-5}) and (\ref{renorm-consist}) without using the exponentiated form of the soft function, which provides a consistency check on our calculation.
The results for the NLO and NNLO renormalized coefficients of Eqs.~(\ref{expo-3}) and (\ref{expo-4}) are
\begin{eqnarray}
s^{(1)}(q^{+},q^{-},b_{\perp},\mu) &=& C_F  \left\{ -\frac{\pi^2}{12} L_{0,0} + L_{0,2} + J_{0}(b_{\perp}\sqrt{q^{+}q^{-}}) L_{1,1} \right\},\nonumber \\
s^{(2)}(q^{+},q^{-},b_{\perp},\mu) &=&
C_F N_F \left\{ -\left(\frac{41}{162}-\frac{5\pi^2}{144}-\frac{5}{36}\zeta(3)\right) L_{0,0} + \left(\frac{7}{27}-\frac{\pi^2}{36}\right) L_{0,1}
\right. \nonumber \\
&& - \frac{5}{18} L_{0,2} + \frac{1}{12} L_{0,3} - \left(\frac{1}{6}T(b_{\perp}\sqrt{q^{+}q^{-}})+\frac{5}{18}J_0(b_{\perp}\sqrt{q^{+}q^{-}})\right) L_{1,1} \nonumber \\
&& \left. + \frac{1}{6}J_0(b_{\perp}\sqrt{q^{+}q^{-}}) L_{0,2} \right\} + \nonumber \\
&& C_F C_A \left\{ \frac{\zeta(3)}{36} L_{0,0} - \frac{11\pi^2}{144} L_{0,1} + \frac{11}{24} L_{0,3} - \frac{11}{12} T(b_{\perp}\sqrt{q^{+}q^{-}}) L_{1,1} \right. \nonumber \\
&& \left. + \frac{1}{12} J_0(b_{\perp}\sqrt{q^{+}q^{-}}) L_{1,2} \right\},
\end{eqnarray}
where $J_0(x)$ is the standard Bessel function. 

\section{Conclusions \label{sec:conc}}

We have described a computation of the exclusive soft function for Drell-Yan production of electroweak gauge bosons through next-to-next-to-leading order in perturbation theory.  This object is required for the resummation 
of low $p_T$ logarithms through next-to-next-to-leading logarithmic accuracy in the SCET approach of Refs.~\cite{Mantry:2009qz, Mantry:2010mk}.  Results for both the anomalous dimension and the finite soft function have 
been presented, and all relevant technical details have been explained.  Adapting these techniques to the computation of the soft function that appears for 
gluon-initiated production of a Higgs boson should be straightforward.  We expect that the exclusive soft function will have further applications in precision studies of differential distributions within SCET, and that our result will be an important step toward enabling these future studies.

We conclude with a few comments on the SCET approach with unintegrated distribution functions to the low $p_T$ distribution, in which the need for this soft function first arose. In the standard TMDPDF approach~\cite{Collins:1999dz, Belitsky:2002sm, Hautmann:2007uw, Collins:1992kk, Ji:2004wu, Ji:2002aa, Cherednikov:2007tw, Cherednikov:2008ua,Cherednikov:2009wk, Aybat:2011zv}, rapidity divergences arise in perturbative computations that require additional regulators beyond the standard dimensional regularization.  The need for similar regulators also arises in SCET approaches to related observables~\cite{Chiu:2011qc}.  In our approach, the Impact-parameter Beam Functions (iBFs) and the Inverse Soft function (iSF) are more differential in momentum coordinates than the corresponding objects in the TMDPDF formalism. As a result, in perturbative computations of the iBFs and the iSF the rapidity divergences are regulated by the physical kinematics of the process.  An investigation of the relationship between the our approach with iBFs and the iSF and the TMDPDF formalism is worth pursuing in future studies. However, some 
recent results~\cite{Becher:2010tm} based on the TMDPDF formalism do not include the soft function needed for a proper treatment of soft radiation and lack operator definitions, preventing any rigorous field-theoretic interpretation and rendering comparison with our results difficult.

We have performed the first step needed for NNLO studies of low $p_T$-transverse momentum distributions, using the SCET approach with unintegrated distributions functions, by computing the relevant exclusive soft function at this order. These results are also the first step towards achieving low-$p_T$ resummation at NNLL accuracy. We look forward to the further development of our approach to low-$p_T$ distributions. 

\section*{Acknowledgements}
This work is supported by the U.S. Department of Energy, Division of High Energy Physics, under contract DE-AC02-06CH11357 and the grants DE-FG02-95ER40896 and DE-FG02-08ER4153, and by Northwestern University.

\appendix

\section{Auxiliary integrals}
\label{integrals}
We compile here several integrals that appear during the course of our calculation.  We define 
\begin{equation}
[dk]=d^{d}k_1 \; d^{d}k_2 \; \delta_{+}(k_1^2) \; \delta_{+}(k_2^2) \; \delta^d(q-k_1-k_2).
\end{equation}
It is straightforward to derive the following by direct integration:
\begin{eqnarray}
\int [dk] &=& \pi^{1-\epsilon} \frac{\Gamma(1-\epsilon)}{\Gamma(1-2\epsilon)} \frac{1}{2(1-2\epsilon)} (q^2)^{-\epsilon}, \\
\int [dk] \frac{1}{k^{+}_1} &=& - \pi^{1-\epsilon} \frac{\Gamma(1-\epsilon)}{\Gamma(1-2\epsilon)} \frac{1}{2\epsilon} \frac{(q^2)^{-\epsilon}}{q^{+}}, \\
\int [dk] \frac{1}{k^{+}_1 k^{-}_1} &=& - \pi^{1-\epsilon} \frac{\Gamma(1-\epsilon)}{\Gamma(1-2\epsilon)} \frac{1}{\epsilon} \frac{(q^{+}q^{-})^{\epsilon}}{(q^2)^{1+2\epsilon}} {_2F_1(-\epsilon,-\epsilon;1-\epsilon;\frac{{q_\perp}^2}{q^{+}q^{-}})}, \\
\int [dk] \frac{1}{k^{+}_1 k^{-}_2} &=& - \pi^{1-\epsilon} \frac{\Gamma(1-\epsilon)}{\Gamma(1-2\epsilon)} \frac{1}{\epsilon} \frac{(q^{+}q^{-})^{\epsilon}}{({q_\perp}^2)^{1+\epsilon}(q^2)^{\epsilon}} {_2F_1(-\epsilon,-\epsilon;1-\epsilon;\frac{q^2}{q^{+}q^{-}})}.
\end{eqnarray}
The following integrals are useful in performing Fourier transformations:
\begin{eqnarray}
\int &d^{d-2}q_{\perp}& e^{-i\vec{q}_{\perp}\cdot\vec{b}_{\perp}} \; q^m \; q_{\perp}^n \; \theta_{+}(q^2) \nonumber \\
&=\pi^{1-\epsilon}& (q^+ q^-)^{1-\epsilon+(m+n)/2}
\frac{\Gamma(1+m/2)\Gamma(1-\epsilon+n/2)}{\Gamma(1-\epsilon)\Gamma(2-\epsilon+(m+n)/2)} \nonumber \\
&& \times {\;_1F_2(1-\epsilon+\frac{n}{2};1-\epsilon,2-\epsilon+\frac{m+n}{2};-\frac{{b_{\perp}}^2 q^{+}q^{-}}{4})}, \\
\int &d^{d-2}q_{\perp}& e^{-i\vec{q}_{\perp}\cdot\vec{b}_{\perp}} \; q^{-2-4\epsilon} \; \theta_{+}(q^2) {\;_2F_1(-\epsilon,-\epsilon;1-\epsilon;\frac{{q_{\perp}}^2}{q^+ q^-})} \nonumber \\
&=\pi^{1-\epsilon}& (q^+ q^-)^{-3\epsilon} \frac{\Gamma(1-\epsilon)}{-2\epsilon\Gamma(1-2\epsilon)}
{\;_0F_1(1-3\epsilon;-\frac{{b_{\perp}}^2 q^{+}q^{-}}{4})} + {\rm O}(\epsilon^2), \\
\int &d^{d-2}q_{\perp}& e^{-i\vec{q}_{\perp}\cdot\vec{b}_{\perp}} \; q^{-2\epsilon} \; q_{\perp}^{-2-2\epsilon} \; \theta_{+}(q^2) {\;_2F_1(-\epsilon,-\epsilon;1-\epsilon;\frac{q^2}{q^+ q^-})} \nonumber \\
&=\pi^{1-\epsilon}& (q^+ q^-)^{-3\epsilon} \frac{\Gamma(1-\epsilon)}{-2\epsilon\Gamma(1-2\epsilon)}
{\;_1F_2(-2\epsilon;1-\epsilon,1-3\epsilon;-\frac{{b_{\perp}}^2 q^{+}q^{-}}{4})} + {\rm O}(\epsilon^2), \\
\int &d^{d-2}q_{\perp}& e^{-i\vec{q}_{\perp}\cdot\vec{b}_{\perp}} \; q^{-2-2\epsilon} \; q_{\perp}^{-2\epsilon} \; \theta_{+}(q^2) {\;_2F_1(-\epsilon,-\epsilon;1-\epsilon;\frac{q^2}{q^+ q^-})} \nonumber \\
&=\pi^{1-\epsilon}& (q^+ q^-)^{-3\epsilon} \frac{\Gamma(1-2\epsilon)}{-\epsilon\Gamma(1-3\epsilon)}
{\;_3F_2(-\epsilon,-\epsilon,-\epsilon;1-\epsilon,1-3\epsilon;1)} \nonumber \\
&& \times {\;_1F_2(1-2\epsilon;1-\epsilon,1-3\epsilon;-\frac{{b_{\perp}}^2 q^{+}q^{-}}{4})} + {\rm O}(\epsilon^2),
\end{eqnarray}
\begin{eqnarray}
&& \int dq^+ dq^- e^{-iq^- b^+/2}e^{-iq^+ b^-/2} (q^+ q^-)^{l} {\;_mF_n(a_1,a_2,\ldots,a_m;b_1,b_2,\ldots,b_n;-\frac{q^+ q^- b_{\perp}^2}{4})} \nonumber \\
&& = \left(-\frac{b^+ b^-}{4}\right)^{-l-1} \Gamma^2(l+1)
{\;_{m+2}F_n(l+1,l+1,a_1,\ldots,a_m;b_1,\ldots,b_n;\frac{b_{\perp}^2}{b^+ b^-})}.
\end{eqnarray}
The $+$ subscript on the delta and step function denotes that only the positive energy solution is taken. Although some integrals are only correct up to ${\rm O}(\epsilon^2)$, they become exact in the limit of $b_{\perp} \rightarrow 0$ or in the presence of $\delta(q^\pm)$.

\section{Hypergeometric expansions}
For completeness, we present here several expansions of hypergeometric functions that we found useful in our analysis.  These can be simply obtained using the series expansion of the hypergeometric function, expanding the resulting  Gamma functions in $\epsilon$, and using known techniques for summing the resulting series in terms of known functions~\cite{Moch:2001zr}.

\begin{eqnarray}
{\;_pF_q(a_1,a_2,\ldots,a_p;b_1,b_2,\ldots,b_q;0)} &=& 1 \\
{\;_2F_1(a,b;c;1)} &=& \frac{\Gamma(c)\Gamma(c-a-b)}{\Gamma(c-a)\Gamma(c-b)} \\
{\;_3F_2(-\epsilon,-\epsilon,-\epsilon;1-\epsilon,1-3\epsilon;1)} &=& 1 - {\rm \zeta}(3)\epsilon^3 - \frac{17\pi^4}{360}\epsilon^4 + {\rm O}({\epsilon}^5) \\
{\;_0F_1(1-a\epsilon;-\frac{z^2}{4})} &=& \Gamma(1-a\epsilon) \left\{ J_0(z) + a\epsilon \left[J_0(z)\ln(\frac{z}{2})-\frac{\pi}{2}Y_0(z)\right] 
\right. \nonumber \\
&& \left. + {\rm O}({\epsilon}^2) \right\} \\
{\;_1F_2(1-2\epsilon;1-\epsilon,1-3\epsilon;-\frac{z^2}{4})} &=& \frac{\Gamma(1-\epsilon)\Gamma(1-3\epsilon)}{\Gamma(1-2\epsilon)}
\left\{ J_0(z) + 2\epsilon \left[J_0(z)\ln(\frac{z}{2})-\frac{\pi}{2}Y_0(z)\right] \right. \nonumber \\
&& \left. + {\rm O}({\epsilon}^2) \right\}
\end{eqnarray}
\begin{eqnarray}
\label{2F1exp1}
{\;_2F_1(-a\epsilon,-a\epsilon;1-a\epsilon;z)} &=& 1 + a^2 \epsilon^2 {\rm Li}_2(z) - a^3 \epsilon^3 \left[{\rm S}_{1,2}(z)-{\rm Li}_3(z)\right]
+ \nonumber \\
&& a^4 \epsilon^4 \left[ {\rm Li}_4(z)+{\rm S}_{1,3}(z)-{\rm S}_{2,2}(z) \right] + {\rm O}({\epsilon}^5) \\
\label{2F1exp2}
{\;_2F_1(-2\epsilon,-2\epsilon;1-3\epsilon;z)} &=& 1 + 4\epsilon^2 {\rm Li}_2(z) - 4\epsilon^3 \left[{\rm S}_{1,2}(z)-3{\rm Li}_3(z)\right] + \nonumber \\
&& 2\epsilon^4 \left[ {\rm Li}_2^2(z)+18{\rm Li}_4(z)+2{\rm S}_{1,3}(z)-10{\rm S}_{2,2}(z) \right] + \nonumber \\
&& {\rm O}({\epsilon}^5) \\
{\;_2F_1(-2\epsilon,-2\epsilon;1-\epsilon;z)} &=& 1 + 4\epsilon^2 {\rm Li}_2(z) - 4\epsilon^3 \left[3{\rm S}_{1,2}(z)-{\rm Li}_3(z)\right] + \nonumber \\
&& 2\epsilon^4 \left[ {\rm Li}_2^2(z)+2{\rm Li}_4(z)+18{\rm S}_{1,3}(z)-10{\rm S}_{2,2}(z) \right] + \nonumber \\
&& {\rm O}({\epsilon}^5) \\
{\;_2F_1(-2\epsilon,-2\epsilon,1-2\epsilon;1-\epsilon,1-3\epsilon;z)} &=& 1 + 4\epsilon^2 {\rm Li}_2(z) - 8\epsilon^3 \left[{\rm S}_{1,2}(z)-{\rm Li}_3(z)\right] + \nonumber \\
&& \epsilon^4 \left[ {\rm Li}_2^2(z)+10{\rm Li}_4(z)+8{\rm S}_{1,3}(z)-12{\rm S}_{2,2}(z) \right]  + \nonumber \\
&& {\rm O}({\epsilon}^5) \\ 
\label{3F2exp}
{\;_3F_2(-2\epsilon,-2\epsilon,-2\epsilon;1-\epsilon,1-3\epsilon;z)} &=& 1 - 8\epsilon^3 {\rm Li}_3(z) - 16\epsilon^4 \left[ 2{\rm Li}_4(z)-{\rm S}_{2,2}(z) \right] + {\rm O}({\epsilon}^5) \nonumber \\
\end{eqnarray}
$J_0(z)$ and $Y_0(z)$ are the standard Bessel functions, and ${\rm Li}_n$ denotes the standard polylogarithmic functions.
In addition, Nielsen's generalized polylogarithms, denoted by  ${\rm S}_{n,p}(z)$, appear.  In the final finite results for the soft function, only two of these functions appear.  They can be 
exchanged for the standard polylogarithms using the following identities:
\begin{eqnarray}
{\rm S}_{1,2}(z) &=& \frac{\ln^2(1-z)\ln(z)}{2} + \ln(1-z){\rm Li}_2(1-z) - {\rm Li}_3(1-z) + \zeta(3), \\
{\rm S}_{1,3}(z) &=& -\frac{\ln^3(1-z)\ln(z)}{6} - \frac{\ln^2(1-z){\rm Li}_2(z)}{2} + \ln(1-z){\rm Li}_3(1-z) - {\rm Li}_4(1-z) \nonumber \\
&& + \frac{\pi^4}{90}.
\end{eqnarray}


\begin{thebibliography}{99}

\bibitem{Dokshitzer:1978yd}
Y.~L. Dokshitzer, D.~Diakonov, and S.~I. Troian,
\newblock Phys. Lett. {\bf B79}, 269 (1978).

\bibitem{Parisi:1979se}
G.~Parisi and R.~Petronzio,
\newblock Nucl. Phys. {\bf B154}, 427 (1979).

\bibitem{Curci:1979bg}
G.~Curci, M.~Greco, and Y.~Srivastava,
\newblock Nucl. Phys. {\bf B159}, 451 (1979).

\bibitem{Collins:1981uk}
J.~C. Collins and D.~E. Soper,
\newblock Nucl. Phys. {\bf B193}, 381 (1981).

\bibitem{Collins:1984kg}
J.~C. Collins, D.~E. Soper, and G.~Sterman,
\newblock Nucl. Phys. {\bf B250}, 199 (1985).

\bibitem{Kauffman:1991jt}
R.~P. Kauffman,
\newblock Phys. Rev. {\bf D44}, 1415 (1991).

\bibitem{Yuan:1991we}
C.~P. Yuan,
\newblock Phys. Lett. {\bf B283}, 395 (1992).

\bibitem{Ellis:1997ii}
R.~K. Ellis and S.~Veseli,
\newblock Nucl. Phys. {\bf B511}, 649 (1998), hep-ph/9706526.

\bibitem{Berger:2002ut}
E.~L. Berger and J.-w. Qiu,
\newblock Phys. Rev. {\bf D67}, 034026 (2003), hep-ph/0210135.

\bibitem{Bozzi:2003jy}
G.~Bozzi, S.~Catani, D.~de~Florian, and M.~Grazzini,
\newblock Phys. Lett. {\bf B564}, 65 (2003), hep-ph/0302104.

\bibitem{Bozzi:2005wk}
G.~Bozzi, S.~Catani, D.~de~Florian, and M.~Grazzini,
\newblock Nucl. Phys. {\bf B737}, 73 (2006), hep-ph/0508068.

\bibitem{Bozzi:2010xn}
G.~Bozzi, S.~Catani, G.~Ferrera, D.~de~Florian, and M.~Grazzini,
\newblock (2010), 1007.2351.


\bibitem{Gao:2005iu}
Y.~Gao, C.~S. Li, and J.~J. Liu,
\newblock Phys. Rev. {\bf D72}, 114020 (2005), hep-ph/0501229.

\bibitem{Idilbi:2005er}
A.~Idilbi, X.-d. Ji, and F.~Yuan,
\newblock Phys. Lett. {\bf B625}, 253 (2005), hep-ph/0507196.

\bibitem{Bauer:2000yr}
C.~W. Bauer, S.~Fleming, D.~Pirjol, and I.~W. Stewart,
\newblock Phys. Rev. {\bf D63}, 114020 (2001), hep-ph/0011336.

\bibitem{Bauer:2001yt}
C.~W. Bauer, D.~Pirjol, and I.~W. Stewart,
\newblock Phys. Rev. {\bf D65}, 054022 (2002), hep-ph/0109045.

\bibitem{Bauer:2002nz}
C.~W. Bauer, S.~Fleming, D.~Pirjol, I.~Z. Rothstein, and I.~W. Stewart,
\newblock Phys. Rev. {\bf D66}, 014017 (2002), hep-ph/0202088.

\bibitem{Mantry:2009qz}
S.~Mantry and F.~Petriello,
\newblock Phys. Rev. {\bf D81}, 093007 (2010).

\bibitem{Mantry:2010mk}
  S.~Mantry and F.~Petriello,
  \newblock Phys.\ Rev.\  D {\bf 83}, 053007 (2011).

\bibitem{Mantry:2010bi}
  S.~Mantry and F.~Petriello,
  \newblock arXiv:1011.0757 [hep-ph].
  
\bibitem{Qiu:2000hf}
  J.~Qiu and X.~Zhang,
   Phys.\ Rev.\  D {\bf 63}, 114011 (2001).

\bibitem{Manohar:2006nz}
  A.~V.~Manohar, I.~W.~Stewart,
  Phys.\ Rev.\  {\bf D76}, 074002 (2007).
  [hep-ph/0605001]

\bibitem{Lee:2006nr}
  C.~Lee and G.~Sterman,
  Phys.\ Rev.\  D {\bf 75}, 014022 (2007).
  [hep-ph/0611061].

\bibitem{Idilbi:2007ff}
  A.~Idilbi and T.~Mehen,
  Phys.\ Rev.\  D {\bf 75}, 114017 (2007).
  [hep-ph/0702022].
  
\bibitem{Collins:1999dz}
  J.~C.~Collins, F.~Hautmann,
  Phys.\ Lett.\  {\bf B472}, 129-134 (2000).
  [hep-ph/9908467].
  
\bibitem{Belitsky:2002sm}
  A.~V.~Belitsky, X.~Ji, F.~Yuan,
  Nucl.\ Phys.\  {\bf B656}, 165-198 (2003).
  [hep-ph/0208038].
  
\bibitem{Hautmann:2007uw}
  F.~Hautmann,
  Phys.\ Lett.\  {\bf B655}, 26-31 (2007).
  [hep-ph/0702196 [HEP-PH]].
  
  \bibitem{Collins:1992kk}
  J.~C.~Collins,
  Nucl.\ Phys.\  {\bf B396}, 161-182 (1993).
  [hep-ph/9208213].
  
\bibitem{Ji:2002aa}
  X.~-d.~Ji, F.~Yuan,
  Phys.\ Lett.\  {\bf B543}, 66-72 (2002).
  [arXiv:hep-ph/0206057 [hep-ph]].
  
  \bibitem{Collins:2003fm}
  J.~C.~Collins,
  Acta Phys.\ Polon.\  {\bf B34}, 3103 (2003).  
  
\bibitem{Ji:2004wu}
  X.~-d.~Ji, J.~-p.~Ma, F.~Yuan,
  Phys.\ Rev.\  {\bf D71}, 034005 (2005).
  [hep-ph/0404183].
  
\bibitem{Cherednikov:2007tw}
  I.~O.~Cherednikov, N.~G.~Stefanis,
  Phys.\ Rev.\  {\bf D77}, 094001 (2008).
  [arXiv:0710.1955 [hep-ph]].
  
\bibitem{Cherednikov:2008ua}
  I.~O.~Cherednikov, N.~G.~Stefanis,
  Nucl.\ Phys.\  {\bf B802}, 146-179 (2008).
  [arXiv:0802.2821 [hep-ph]].
  
\bibitem{Cherednikov:2009wk}
  I.~O.~Cherednikov, N.~G.~Stefanis,
  Phys.\ Rev.\  {\bf D80}, 054008 (2009).
  [arXiv:0904.2727 [hep-ph]].
  
  
\bibitem{Aybat:2011zv}
  S.~M.~Aybat, T.~C.~Rogers,
  [arXiv:1101.5057 [hep-ph]].



\bibitem{Fleming:2006cd}
  S.~Fleming, A.~K.~Leibovich, T.~Mehen,
  Phys.\ Rev.\  {\bf D74}, 114004 (2006).
  [hep-ph/0607121]

\bibitem{Stewart:2009yx}
  I.~W.~Stewart, F.~J.~Tackmann, W.~J.~Waalewijn,
  Phys.\ Rev.\  {\bf D81}, 094035 (2010).

\bibitem{Hoang:2008fs}
  A.~H.~Hoang, S.~Kluth,
  [arXiv:0806.3852 [hep-ph]].
  
\bibitem{Kelley:2011ng}
  R.~Kelley, R.~M.~Schabinger, M.~D.~Schwartz, H.~X.~Zhu,
  [arXiv:1105.3676 [hep-ph]].
  
\bibitem{Monni:2011gb}
  P.~F.~Monni, T.~Gehrmann, G.~Luisoni,
  [arXiv:1105.4560 [hep-ph]].
  
\bibitem{Hornig:2011iu}
  A.~Hornig, C.~Lee, I.~W.~Stewart, J.~R.~Walsh and S.~Zuberi,
  arXiv:1105.4628 [hep-ph].

\bibitem{Bauer:2002ie}
  C.~W.~Bauer, A.~V.~Manohar, M.~B.~Wise,
  Phys.\ Rev.\ Lett.\  {\bf 91}, 122001 (2003).
  [hep-ph/0212255].
  
\bibitem{Bauer:2003di}
  C.~W.~Bauer, C.~Lee, A.~V.~Manohar, M.~B.~Wise,
  Phys.\ Rev.\  {\bf D70}, 034014 (2004).
  [hep-ph/0309278].
  
\bibitem{Fleming:2007xt}
  S.~Fleming, A.~H.~Hoang, S.~Mantry, I.~W.~Stewart,
  Phys.\ Rev.\  {\bf D77}, 114003 (2008).
  [arXiv:0711.2079 [hep-ph]].
  
\bibitem{Fleming:2007qr}
  S.~Fleming, A.~H.~Hoang, S.~Mantry, I.~W.~Stewart,
  Phys.\ Rev.\  {\bf D77}, 074010 (2008).
  [hep-ph/0703207].
 
\bibitem{Catani:1992ua}
  S.~Catani, L.~Trentadue, G.~Turnock, B.~R.~Webber,
  Nucl.\ Phys.\  {\bf B407}, 3-42 (1993).

\bibitem{Korchemsky:1994is}
  G.~P.~Korchemsky, G.~F.~Sterman,
  Nucl.\ Phys.\  {\bf B437}, 415-432 (1995).
  [hep-ph/9411211].

\bibitem{Kidonakis:1998bk}
  N.~Kidonakis, G.~Oderda, G.~F.~Sterman,
  Nucl.\ Phys.\  {\bf B525}, 299-332 (1998).
  [hep-ph/9801268].

\bibitem{Korchemsky:2000kp}
  G.~P.~Korchemsky, S.~Tafat,
  JHEP {\bf 0010}, 010 (2000).
  [hep-ph/0007005].

\bibitem{Schwartz:2007ib}
  M.~D.~Schwartz,
  Phys.\ Rev.\  {\bf D77}, 014026 (2008).
  [arXiv:0709.2709 [hep-ph]].

\bibitem{Hoang:2007vb}
  A.~H.~Hoang, I.~W.~Stewart,
  Phys.\ Lett.\  {\bf B660}, 483-493 (2008).
  [arXiv:0709.3519 [hep-ph]].

\bibitem{Idilbi:2010im}
  A.~Idilbi and I.~Scimemi,
  Phys.\ Lett.\  B {\bf 695}, 463 (2011)
  [arXiv:1009.2776 [hep-ph]].

\bibitem{GarciaEchevarria:2011md}
  M.~Garcia-Echevarria, A.~Idilbi and I.~Scimemi,
  arXiv:1104.0686 [hep-ph].
  
\bibitem{Belitsky:1998tc}
  A.~V.~Belitsky,
  Phys.\ Lett.\  {\bf B442}, 307-314 (1998).
  
  \bibitem{Gatheral:1983cz}
  J.~G.~M.~Gatheral,
  Phys.\ Lett.\  {\bf B133}, 90 (1983).

\bibitem{Frenkel:1984pz}
  J.~Frenkel, J.~C.~Taylor,
  Nucl.\ Phys.\  {\bf B246}, 231 (1984).

\bibitem{Chiu:2011qc}
  J.~Y.~Chiu, A.~Jain, D.~Neill and I.~Z.~Rothstein,
  arXiv:1104.0881 [hep-ph].

\bibitem{Becher:2010tm}
  T.~Becher, M.~Neubert,
  [arXiv:1007.4005 [hep-ph]].

\bibitem{Moch:2001zr}
  S.~Moch, P.~Uwer, S.~Weinzierl,
  J.\ Math.\ Phys.\  {\bf 43}, 3363-3386 (2002).

\end{thebibliography}
\end{document}